\documentstyle[12pt]{article}
\begin{document}
\title{On the symmetries of the 1+1 dimensional relativistic fluid dynamics}
\author{C Alexa \\ 
I.F.I.N. - High Energy Dept. Magurele-Bucharest 76900 \\
\and
D Vrinceanu \\
University of Bucharest, Dept. of Theoretical Physics, Bucharest}
\maketitle
\begin{abstract}
A very intuitive description of nucleus-nucleus collision phenomena is 
provided by the relativistic fluid dynamics. We consider a 1+1 dimensional 
relativistic
imperfect fluid flow to approximate the high energy heavy ion collision. 
The article investigates the application of the continuous symmetry group 
on the relativistic fluid energy-momentum tensor conservation equations 
in the ultrarelativistic limit $\gamma \rightarrow \infty$.
\end{abstract}
\section{Introduction}
There are three important theoretical approaches of the heavy ion collisions: 
fluid dynamics, Boltzman equation and statistical models.

A simple but intuitive description of the general nuclear phenomena is given
by fluid dynamics and it has been successfully applied to the
ultrarelativistic central heavy ion collisions \cite{bjo}.

The Boltzman equation has already been applied to heavy ion collisions for
relative low energies \cite{sch} and there is also a relativistic Boltzmann
equation using J\"uttner distribution $f^{Juttner}= 1/( 2\pi \hbar)^3\;
\exp \left((\mu -p^\alpha u_\alpha )/T\right) $where $\mu $
is the chemical potential, $u_\alpha $ and $p^\alpha $ are the four-velocity
and four-momentum of the particle, $T$ is the temperature parameter \cite
{cser94}.

In statistical models, one use the equilibrium thermodynamics to get the
properties of the system. One of the main interest is the study of the phase
transition from hadrons to quarks and gluons. At this point we can mention
that the lattice quantum chromodynamics and subsequent use of Monte Carlo
techniques enabled us to study physical observables over the entire
temperature range from $0$ to $\infty $ \cite{cle} and the very existence of
a phase transition.

A consistent way of investigation of the high energy heavy ion collisions is
the manifest covariant formulation of relativistic fluid dynamics. An
excellent review about the status of fluid dynamics models for relativistic
heavy ion reactions is the article written by D.Strottman \cite{str}. One of
the arguments to justify relativistic fluid dynamics is that the mean free
path $\Lambda $ is smaller than the size $L$ of the nucleus. A simple
estimate of the order of magnitude of $\Lambda $ is about $1\,\,fm,$ which
is to be compared to the size of heavy nucleus 15 $fm$ (for uranium). On the
other hand, it was showed \cite{mar} that a single particle quantum
mechanics is formally analogous to the Euler hydrodynamics.

Solving the hydrodynamics equations is hardly discouraged by the puzzle of
choosing the rest frame. This situation leads us to acausal and instable
solutions \cite{str}. Furthermore, the choose of the initial conditions is
rather unclear.  In the same time the numerical approaches do not give us
satisfactory quantitative results.

Complementary information may be achieved by exploiting the Lie symmetry
group of the covariant relativistic hydrodynamics equations. The power of
this technique consists in the possibility to explore the properties of
physical systems, like the symmetry structure and the invariants, without
solving the corresponding differential equations. The group structure and
the invariants can help us to reduce the order of the equation and even to
integrate them. Complex systems were successfully studied using this
approach \cite{olver}. In \cite{ale96} we study a simple form of
energy-momentum tensor conservation, where we already obtained symmetries
and invariants.

\section{Energy-momentum tensor conservation equation}

Relativistic fluid dynamics is well describe by the number of particles N
and energy-momentum tensor $T_{\alpha \beta}$ conservation equations \cite
{wei}. In the ideal case we have : 
\begin{eqnarray*}
\begin{array}{lcl}
T_{\alpha \beta }&=&p\eta _{\alpha \beta }+(p+\rho )U_\alpha U_\beta  \\
[2mm]
N_{\alpha} &=& n U_{\alpha} 
\end{array}
\end{eqnarray*}
where $p$ is the pressure, $\rho $ is the energy density, n is the number of
particles density and $U_\alpha :(\gamma \vec \beta, \gamma )$ is the
4-velocity field.

There are two ways of choosing the rest frame : in the Landau way $U_\alpha $
is the energy transport velocity where $T_{i0}=0$ in the rest frame, while
in the Eckart way $U_\alpha $ is the particle transport velocity where $%
N_i=0 $ in the rest frame. The dissipation contribution is introduced by
redefining the energy-momentum and number of particle tensor by adding
correction terms : 
\begin{eqnarray*}
\begin{array}{lcl}
T_{\alpha \beta }&=&p\eta _{\alpha \beta }+(p+\rho )U_\alpha U_\beta +\Delta
T_{\alpha \beta }, \\
[2mm]
N_\alpha &=&nU_\alpha +\Delta N_\alpha 
\end{array}
\end{eqnarray*}
In the Eckart frame $\Delta N_\alpha =0$, so the dissipation contribution is
present only in the energy-momentum terms. In the following we choose the
Eckart approach. The construction of the most general dissipation term $%
\Delta T_{\alpha \beta }$ is coming up from the positivity of the entropy
production: 
$$
\Delta T^{\alpha \beta }=-\eta H^{\alpha \gamma }H^{\beta \delta }W_{\gamma
\delta }-\chi \left( H^{\alpha \gamma }U^\beta -H^{\beta \gamma }U^\alpha
\right) Q_\gamma -\zeta H^{\alpha \beta }\partial _\gamma U^\gamma 
$$
where we have shear tensor: 
$$
W_{\alpha \beta }=\partial _\beta U_\alpha +\partial _\alpha U_\beta -\frac 2%
3\eta _{\alpha \beta }\,\partial _\gamma U^\gamma , 
$$
heat-flow vector: 
$$
Q_\alpha =\partial _\alpha T+T\,\,U^\beta \,\partial _\beta \,U_\alpha , 
$$
T is the temperature and projection tensor on the hyperplane normal to $%
U_\alpha $ 
$$
H_{\alpha \beta }=\eta _{\alpha \beta }+U_\alpha \,U_\beta \,\,. 
$$
We may identify $\chi ,\eta ,\zeta $ as the coefficients of heat conduction,
shear viscosity and bulk viscosity.

Making some calculations we can write the energy-momentum tensor in the
following form : 
\begin{eqnarray*}
T_{\alpha \beta } & = & p\eta _{\alpha \beta }+(p+\rho )U_\alpha \,U_\beta \\
[2mm]
& & -\eta \left[ \partial _\alpha U_\beta +\partial _\beta U_\alpha -\frac
23(\eta _{\alpha \beta }+U_\alpha \,U_\beta )\,\partial _\gamma U^\gamma
+\,U^\gamma \partial _\gamma (U_\alpha \,U_\beta )\right]  \\ 
[2mm]
& &-\chi \left[ \,U_\alpha \,\partial _\beta \,T+\,U_\beta \,\partial _\alpha
\,T+2U_\alpha \,U_\beta U^\gamma \partial _\gamma T+TU^\gamma \partial
_\gamma (U_\alpha \,U_\beta )\right]  \\ 
[2mm]
& &-\zeta \left( \eta _{\alpha \beta }+U_\alpha \,U_\beta \right) \partial
_\gamma U^\gamma 
\end{eqnarray*}

With this form we rewrite the energy-momentum conservation :
\begin{eqnarray*}
\partial ^\alpha T_{\alpha \beta }&=&\partial ^\beta \left[ p+\left( \frac
23\eta -\zeta \right) \partial ^\alpha U_\alpha \right] \\ 
[2mm]
& &+\partial ^\alpha \left\{ U_\alpha \,U_\beta \left[ p+\rho +\left( \frac
23\eta -\zeta \right) \partial _\gamma U^\gamma -2\chi U^\gamma \partial
_\gamma T\right] \right\} \\ 
[2mm]
& &-\partial ^\alpha \left[ \left( \eta +\chi T\right) U^\gamma \partial
_\gamma (U_\alpha \,U_\beta )-\eta \left( \partial _\alpha U_\beta +\partial
_\beta U_\alpha \right) -\chi \left( U_\alpha \,\partial _\beta
\,T+\,U_\beta \,\partial _\alpha \,T\right) \right] 
\end{eqnarray*}

We can see this equation as a polinom in $U$, with the power $3,2,1,0$ .
From this expression we select only the highest power of $U$ and the free
terms. Because $U$ is proportional with $\gamma $ Lorentz $U^3\gg U^2\gg
U\gg 1$ in the ultrarelativistic limit, or we can make a boost to a system
were $\gamma $ is big enough to obtain to above inequality. This
approximation is convenable for us because the form of the equation is much
more simpler. In the same time we don't have any physical contradictions -
as we will see later on Lorentz transformation will be one of the symmetry
group transformation - and we don't loose physical information about the
fluid.

Doing this one can find the following form of the equation : 
$$
\partial ^\alpha T_{\alpha \beta }=\partial _\beta \left[ p+\left( \frac 23%
\eta -\zeta \right) \partial ^\alpha U_\alpha \right] -2\chi \cdot \partial
^\alpha \left( U_\alpha \,U_\beta U^\gamma \partial _\gamma T\right) 
$$

In the final form we have in the 1+1 dimensional approximation two equations: 
\begin{eqnarray*}
\begin{array}{lcl}
p_x+\left( \frac 23\eta -\zeta \right) \left( u_{xx}-v_{xt}\right) -2\chi
\left( u^3T_{xx}+uv^2T_{tt}-2u^2vT_{xt}\right) &=& 0 \\ 
[2mm]
p_t+\left( \frac 23\eta -\zeta \right) \left( u_{xt}-v_{tt}\right) -2\chi
\left( u^2vT_{xx}+v^3T_{tt}-2uv^2T_{xt}\right) &=& 0 
\end{array}
\end{eqnarray*}
where $p_x,p_t,$ etc. means the partial derivative of $p$ with respect to 
$x,t,$ etc. The $u$ and $v$ are the spatial and temporal components of the
velocity field $U\left( \gamma \vec v,\gamma \right) =U\left( u,v\right)$.

\section{Symmetry group of differential equations}

The symmetry group of a system of differential equations is the largest
local group of transformation acting on the independent and dependent
variables of the system with the property that it transform solutions of the
system to other solutions.

We restrict our attention to local Lie group of symmetries, leaving aside
problems involving discrete symmetries such as reflections.

Let ${\cal S}$ be a system of differential equations. A symmetry-group of
the system ${\cal S}$ is a local group of transformations G acting on an
open subset M of the space of independent and dependent variables for the
system with the property that whenever u=f(x) is a solution of ${\cal S}$,
and whenever $g\cdot f$ is defined for $g\in G$, then $u=g\cdot f(x)$ is
also a solution of the system.

Applying the standard procedure \cite{olver} we solve the defining equations
for the symmetry group of the given system of differential equations. We
find a nine parameter symmetry group for the ultrarelativistic fluid
dynamics equations. The basis of the corresponding solvable Lie algebra of
this group is : 
\begin{eqnarray*} 
\begin{array}{ccll}
V_1 &=& \partial _x 
&\mbox{(spatial translation) } \\
[2mm]
V_2 &=& \partial _t 
&\mbox{(temporal translation)} \\
[2mm]
V_3 &=& \partial_T
&\mbox{(temperature translation)} \\
[2mm]
V_4 &=& \partial_p
&\mbox{(presure translation)} \\
[2mm]
V_5 &=& x \partial_T
&\mbox{} \\
[2mm]
V_6 &=& t \partial_T
&\mbox{} \\
[2mm]
V_7 &=& u\, \partial _u + v\, \partial _v - 2 T\, \partial _T - p\, \partial_p  
&\mbox{(dilatations)} \\
[2mm]
V_8 &=&  t\, \partial _x  +  x\, \partial _t  - u\, \partial _v - v\, \partial _u
&\mbox{(hiperbolic rotations)} \\
[2mm]
V_9 &=& 4x\, \partial _x  +  4t\, \partial _t  - u\, \partial _u - v\, \partial _v+
             2T\,\partial_T +5p\,\partial_p
&\mbox{(dilatations)} 
\end{array}
\end{eqnarray*}
The symmetry group infinitesimal generator is defined by : 
$$
\vec {{\cal V}}=\xi \partial _x+\tau \partial _t+\Phi \partial _u+\Psi
\partial _v+\Gamma \partial _T+\Omega \partial _p 
$$
where 
\begin{eqnarray*}
\xi &=& c_1+c_8\cdot t+4c_9\cdot x \\
[2mm]
\tau &=& c_2+c_8\cdot x+4c_9\cdot t \\
[2mm]
\Phi &=& (c_7-c_9)u-c_8v \\
[2mm]
\Psi &=& (c_7-c_9)v-c_8u \\
[2mm]
\Gamma &=& c_3+c_5x+c_6t-2(c_7-c_9)T \\
[2mm]
\Omega &=& c_4-(c_7-5c_9)p 
\end{eqnarray*}
and $a_i$ are arbitrary group parameters.

We calculate the second order prolongation of $\vec {{\cal V}}$ : 
\begin{eqnarray*}
pr^{(2)}\vec {{\cal V}}
&=&\xi \partial _x+\tau \partial _t+\Phi \partial _u+\Psi \partial
_v+\Gamma \partial _T+\Omega \partial _p \\ 
[2mm]
& &+\Phi ^x\partial _{u_x}+\Phi
^t\partial _{u_t}+\Psi ^x\partial _{v_x}+\Psi ^t\partial _{v_t}+\Gamma
^x\partial _{T_x}+\Gamma ^t\partial _{T_t}+\Omega ^x\partial _{p_x}+\Omega
^t\partial _{p_t} \\ 
[2mm]
& &+\Phi ^{xx}\partial _{u_{xx}}+\Phi ^{xt}\partial _{u_{xt}}+\Phi
^{tt}\partial _{u_{tt}}+\Psi ^{xx}\partial _{v_{xx}}+\Psi ^{xt}\partial
_{v_{xt}}+\Psi ^{tt}\partial _{v_{tt}} \\ 
[2mm]
& &+\Gamma ^{xx}\partial _{T_{xx}}+\Gamma ^{xt}\partial _{T_{xt}}+\Gamma
^{tt}\partial _{T_{tt}}+\Omega ^{xx}\partial _{p_{xx}}+\Omega ^{xt}\partial
_{p_{xt}}+\Omega ^{tt}\partial _{p_{tt}} 
\end{eqnarray*}
The coefficient functions of the prolongation of $pr^{\left( n\right) }\vec 
{{\cal V}}$ are given by the following formula \cite{olver} :
$$
\Phi _\alpha ^{J,k}=D_k\Phi _\alpha ^J-\sum\limits_{i=1}^pD_k\xi
^iu_{J,i}^\alpha 
$$
where $p$ is the number of the independent variables, $\xi ^i$ are the
coefficients of the partial derivative of the independent variables $(x,t)$
- here $\xi ^i=(\xi ,\tau )$ , $q$ is the number of dependent variables $%
u=(u,v,T,p)$ - in this case we have $q=4$ so $\Phi _\alpha =\left( \Phi
,\Psi ,\Gamma ,\Omega \right) $ and $D_i$ is the total derivative 
$$
D_if=\frac{\partial f}{\partial x^i}+\sum\limits_{\alpha
=1}^q\sum\limits_Ju_{J,i}^\alpha \frac{\partial f}{\partial u_J^\alpha } 
$$
$$
u_{J,i}^\alpha =\frac{\partial u_J^\alpha }{\partial x^i}=\frac{\partial
^{k+1}u^\alpha }{\partial x^i\partial x^{j_1}...\partial x^{j_k}} 
$$
$J=(j_1,...,j_k),0\leq \# J\leq n$ , $n$ is the highest order derivative
appearing in $f$. For example $\Phi ^x=D_x(\Phi -\xi u_x-\tau u_t)+\xi
u_{xx}+\tau u_{xt}$ , $\Psi ^{xx}=D_x^2(\Psi -\xi v_x-\tau v_t)+\xi
v_{xxx}+\tau v_{xxt}$ and $D_x\Gamma =\Gamma _x+\Gamma _uu_x+\Gamma
_vv_x+\Gamma _TT_x+\Gamma _pp_x$.

An n-th order differential invariant of a group G is a smooth function
depending on the independent and dependent variables and their derivatives,
invariant on the action of the corresponding n-th prolongation of G \cite
{olver}.

We use the method of characteristics for computing the invariants for 
the prolongation of the every element of the Lie algebra ;
then we re-express the next vector in terms of first vector's invariants; we
repet this procedure until we obtaine all global invariants of the vectors. 
In the end we obtained seven global independent invariants of the symmetry
group: 
\begin{eqnarray*}
I_1&=&\left( u+v\right) \left( p_x+p_t\right) \exp \left( -\frac{2v_x}{u_x+v_t}%
\right) \\  
[2mm]
I_2&=&\sqrt{\left( u^2-v^2\right) \left( p_x^2-p_t^2\right) }\frac{u_t-v_x}{u_x-v_t} \\
[2mm]
I_3&=&\sqrt{p_x^2-p_t^2}\frac{\left( p_x+p_t\right) \left( u_t-v_x\right) }{p_{xx}-p_{tt}}\exp \left( -\frac{2v_x}{u_x+v_t}\right) \\
[2mm]
I_4&=&\frac{\left( p_{xx}-p_{tt}\right) ^2\left( u+v\right) ^2}{\left(
p_x+p_t\right) ^2\left( u_xv_t-u_tv_x\right) } \\
[2mm]
I_5&=&\frac{\left( u_xv_t-u_tv_x\right) \left( p_x+p_t\right) ^2}{\left(
u+v\right) ^2\left( T_{xx}-T_{tt}\right) } \\
[2mm]
I_6&=&\frac{\left( u_xv_t-u_tv_x\right) ^3\left( T_{xx}-T_{tt}\right) ^3}{\left( p_{xt}^2-\frac 12p_{xx}^2-p_{xx}p_{tt}\right) ^2}\left[
u_{xt}^2+u_{xt}v_{xx}-\frac 12u_{xx}^2-u_{xx}\left( u_{tt}+v_{xx}\right)
\right] ^{-2} \\
[2mm]
I_7&=&\frac{\left( p_{xt}^2-\frac 12p_{xx}^2-p_{xx}p_{tt}\right) \left(
T_{xt}^2-\frac 12T_{xx}^2-T_{xx}T_{tt}\right) }{\left[
v_{xt}^2+v_{xt}u_{xx}-\frac 12v_{xx}^2-v_{xx}\left( u_{xt}+v_{tt}\right)
\right] ^2}\left[ u_{xt}^2+u_{xt}v_{xx}-\frac 12u_{xx}^2-u_{xx}\left(
u_{tt}+v_{xx}\right) \right]  
\end{eqnarray*}

\section{Conclusions}

Relativistic imperfect fluid flow seems to be a very good approach of the
ultrarelativistic heavy ion collision, because we can very well suppose a
zero mean free path and a instantaneous local equilibrium.

In spite of the simplification of the equations, our results indicate a
nontrivial structure of the symmetry group for the 1+1 dimensional
relativistic fluid dynamics equations. 

As expected due to the covariant formulation of the theory, the symmetry 
group contain the Poincare generators. And of course, other system 
specific transformations mentioned above are present.

Beside the physical meaning that can be associated to the invariants, they
can be used for reducing the order of the original equations. Doing this one
can hope to find simpler equations that can be integrated; unfortunately
only for special type of groups one can find the general solutions of the
equations by quadratures alone.

\end{document}